\documentclass[journal]{IEEEtran}
\usepackage[cmex10]{mathtools}
\usepackage{amsfonts}
\usepackage[numbers, square, comma, sort&compress]{natbib}
\usepackage{algorithm}
\usepackage{algorithmic}
\usepackage{url,color}

\newtheorem{theorem}{Theorem}

\newcommand{\st}[0]{\text{s.t.}}
\newcommand{\trace}[1]{\operatorname{Trace}\left(#1\right)}
\newcommand{\rank}[1]{\operatorname{Rank}\left(#1\right)}

\newcommand{\sign}[1]{\operatorname{Sign}\left(#1\right)}

\newcommand{\prob}[1]{\mathbb{P}\left(#1\right)}

\newcommand{\bR}[0]{\mathbb{R}}
\newcommand{\bC}[0]{\mathbb{C}}

\newcommand{\bS}[0]{\mathbb{S}}

\newcommand{\cN}[0]{\mathcal{N}}
\newcommand{\cM}[0]{\mathcal{M}}
\newcommand{\cF}[0]{\mathcal{F}}

\newcommand{\cO}[0]{\mathcal{O}}
\newcommand{\cP}[0]{\mathcal{P}}
\newcommand{\cS}[0]{\mathcal{S}}
\newcommand{\vc}[0]{\mathbf{c}}
\newcommand{\vf}[0]{\mathbf{f}}
\newcommand{\vg}[0]{\mathbf{g}}
\newcommand{\vs}[0]{\mathbf{s}}

\newcommand{\ve}[0]{\mathbf{e}}
\newcommand{\vu}[0]{\mathbf{u}}
\newcommand{\vv}[0]{\mathbf{v}}
\newcommand{\vw}[0]{\mathbf{w}}
\newcommand{\vx}[0]{\mathbf{x}}
\newcommand{\vy}[0]{\mathbf{y}}
\newcommand{\mf}[0]{\mathbf{F}}

\newcommand{\mh}[0]{\mathbf{H}}
\newcommand{\mm}[0]{\mathbf{M}}
\newcommand{\mq}[0]{\mathbf{Q}}
\newcommand{\ms}[0]{\mathbf{S}}
\newcommand{\mt}[0]{\mathbf{T}}
\newcommand{\mi}[0]{\mathbf{I}}
\newcommand{\mv}[0]{\mathbf{V}}

\newcommand{\hs}[0]{\hat{\vs}}
\newcommand{\ts}[0]{\tilde{\vs}}
\newcommand{\hT}[0]{\widehat{\mt}}

\title{Binary Sequence Set Design for Interferer Rejection\\ in Multi-Branch Modulation}

\author{Dian Mo,~\IEEEmembership{Student~Member,~IEEE}, and Marco F. Duarte,~\IEEEmembership{Senior~Member,~IEEE}
\thanks{This work is partially supported by the National Science Foundation under grant AST-1547278. Portions of this work were presented at IEEE Statistical Signal Processing Workshop~\cite{Mo:2018}.}
\thanks{The authors are with the Department of Electrical and Computer Engineering, University of Massachusetts Amherst, Amherst, MA 01003. Email: mo@umass.edu,
mduarte@ecs.umass.edu.}}

\begin{document}
\maketitle

\begin{abstract}
Wideband communication is often expected to deal with a very wide spectrum, which in many environments of interest includes strong interferers. Thus receivers for the wideband communication systems often need to mitigate interferers present in the received signal to reduce the distortion caused by the amplifier nonlinearity and noise. Recently, a new architecture for communication receivers known as random modulation mixes a signal with different pseudorandom sequences using multiple branches of channels before sampling. While random modulation is used in these receivers to acquire the signal at low sampling rates, the used modulation sequences cannot suppress interferers due to their flat spectra. In previous work, we introduced the design of a single spectrally shaped binary sequence that mitigates interferers to replace the pseudorandom sequence in a channel. However, the designed sequences cannot provide a stable recovery achieved by pseudorandom sequence approaches. In this paper, we extend our previous sequence design to guarantee stable recovery by designing a set of spectrally shaped sequences to be orthogonal to each other. We show that it is difficult to find the necessary number of sequences (equal to the sequence length) featuring mutual orthogonality and introduce oversampling to the sequence set design to improve the recovery performance of the designed sequences. We propose an algorithm for multi-branch sequence design as a binary optimization problem, which is solved using a semidefinite program relaxation and randomized projection. In addition, while it is common to model narrowband interferers as a subspace spanned by a subset of elements from the Fourier basis, we show that the Slepian basis provides an alternative and more suitable compact representation for signals with components contained in narrow spectrum bands. Numerical experiments using the proposed sequence sets show their advantages against pseudorandom sequences and our previous work.
\end{abstract}


\section{Introduction}\label{section:intro}

Receivers for emerging wireless communication systems are expected to deal with a very wide spectrum and adaptively choose which parts of it to extract. The intense demand on the available spectrum for commercial users forces many devices to share the spectrum~\cite{Salzman2001Proceedings-of-, rowe2014spectrally}. As a result, wideband communication systems have caused interference or applications using overlapping regions~\cite{Aubry2014Radar-waveform-, Aubry2015A-new-radar-wav}. Thus, a major issue for wideband communication receivers is to process spectra having very weak signals from a distant source mixed with strong interferers from nearby sources. The presence of strong interferers in the received signals can cause the receiver hardware to operate outside of its linear range, which makes the separation or mitigation of such interferers a key barrier for wideband communication systems since the resulting distortion can be large enough to mask the weaker signals.

Recently, random sequences for wideband signal modulation have been employed in the realization of communication system receivers~\cite{laska2007theory, Mishali2009Blind-Multiband, tropp2010beyond, Mishali:2010}. In essence, a signal is modulated by a pseudorandom sequence and sampled at a sampling rate far below its Nyquist rate at each channel. The random modulation and subsampling alias the spectrum of the signal into baseband so that the output of each channel contains a baseband that is the linear combination of small bands of the input spectrum, where the combination coefficients are proportional to the corresponding spectrum magnitude of the used pseudorandom sequence. Furthermore, the use of multiple instances of those modulate-and-subsample paths (known as branches in the literature) that rely on independent pseudorandom modulation sequences allows for further reduction in the sampling rates required for each one of the branches. Thus, using a large enough number of modulation branches with different pseudorandom sequences allows for successful recovery of the wideband signal, where the number of necessary branches is determined by the occupancy of the spectrum and the sampling rate of each branch~\cite{Mishali:2010}. The resulting multi-branch modulation system can be abstracted as a bank of all-pass filters that preserves all frequency components of interest from the input signal. When the locations of one or more interferers are known, it is desirable to design the modulation sequences to mitigate the interferer, which can provide significant benefit in preserving the receiver hardware within its linear operation conditions and reducing the distortion due to the nonlinearity. Therefore, it is promising to replace the \emph{pseudorandom binary sequence} (PRBS) with a spectrally shaped sequence that effectively implements a notch filter to suppress interferers.

In addition to the strong interferer case described earlier, a similar problem arises in \emph{dynamic spectrum management} (DSM), an approach that allows for flexibility in spectrum use. DSM attempts to determine the frequencies being used or licensed by other applications and selects an optimal subset from the remaining frequencies for use by new or unlicensed applications. The signals for unlicensed applications should be optimized to minimize the interference with licensed signals while keeping their capacities. More specifically, for Direct Sequence Spread Spectrum, a designed spreading code with particular spectral characteristics has been used to shape the unlicensed power spectra~\cite{clancy2006spectrum}. Another similar problem arises in active sensing, which obtains valuable information of targets or the propagation medium by sending probing waveforms toward an area of interest~\cite{Li2009MIMO-Radar-Sign, Zhao2013Enhanced-multis, Liang2014On-Designing-th}. A well-designed waveform is crucial to the performance of active sensing.

In previous work, we presented an algorithm to design a single binary sequence targeted to meet a specific spectrum shape~\cite{Mo2015Design-of-Spect, Mo:2018a}. Such sequences provide a passband and notch for the message and interferer bands, respectively. To maximize the sequence power in the message band while the sequence power in the interferer is upper bounded, the sequence design is formulated as a \emph{quadratically constrained quadratic program} (QCQP). Though solving a QCQP is usually NP-hard~\cite{luo2007approximation}, a method based on a \emph{semidefinite program} (SDP) relaxation and randomized projection can approximately solve the QCQP with good approximation performance. Our approach was inspired by the work of Goemans and Williamson to address the maximum cut problem~\cite{goemans1995improved}, which has been extended to obtain waveforms with a unit module in active sensing~\cite{Maio2008Code-Design-to-, Maio2009Design-of-Phase, Maio2011Design-of-Optim, Cui2014MIMO-Radar-Wave, Aubry2016Forcing-Multipl}. These previous results can be leveraged in multi-branch modulation by mixing the signal with the same designed sequence but with different delays in different branches. The resulting linear measurement operator can be represented by a Toeplitz matrix of which each row is a circularly shifted version of the sequence. However, there are no guarantees that such modulation can provide stable recovery, as shown in the numerical experiments in the sequel.

Our single sequence design approach also selects the Fourier basis to represent a subspace approximation of the interferer band. However, the Fourier basis suffers from spectrum leakage since not all the energy of a signal is preserved if the component frequencies of the input signal are not included in the frequency set sampled by the Fourier basis, which is known as on-grid frequencies. This will cause significant distortion since the sequence designs based on a Fourier subspace characterization have suboptimal performance in mitigating or suppressing interferers that are bandlimited to a narrow spectral band. Alternatively, the Slepian basis can capture almost all the energy for any signal whose component frequencies are contained in a narrow spectral band, regardless of whether they are on-grid or not~\cite{Slepian1976On-bandwidth, Slepian1983Some-Comments-o, Karnik2016The-fast-Slepia, Karnik2016Fast-computatio}.

In this paper, we propose an algorithm to design binary sequence sets for multi-branch modulation that are capable of mitigating strong narrowband interferers. The algorithm is based on our previously proposed single sequence design, which is used repeatedly to obtain spectrally shaped sequences for all branches iteratively. Furthermore, in order to provide stable recovery performance, an additional constraint is included to require the expected sequence in each branch to be approximately orthogonal to all previously designed sequences for other branches. Our main contributions can be detailed as follows. First, we propose a multi-branch sequence set design approach based on our previously introduced single sequence design. Second, we introduce the inner product constraints to obtain sequences that are as mutually orthogonal to each other as possible. Third, we provide an analysis of the connection between the condition number of the measurement operator matrix, which serves as a measure of the stability of the recovery to noise, and the tolerance of the sequence orthogonality, and show that the number of sequences that can be obtained meeting the constraints is dependent on the sequence length. Fourth, we show the necessity of oversampling in the multi-branch sequence design in order to obtain sequences with stable invertibility performance. Finally, we present numerical results to show the advantages of the sequences obtained from the proposed algorithm in interferer mitigation and recovery stability against the PRBS and the sequence from the single sequence design.

This paper is organized as follows. In Section~\ref{section:background}, we review our previous work on single sequence design via SDP relaxation and randomized projection; we also cover some basic information on the Slepian basis needed for sequence set design. In Section~\ref{section:design}, we introduce our algorithm to design sequence sets for multi-branch modulation; furthermore, we present the relationship between the condition number and the orthogonality tolerance and show the necessity of oversampling for the sequence set design. In Section~\ref{section:results}, we present numerical simulations to validate the analysis in Section~\ref{section:design}. Finally, we provide a discussion and conclusions in Section~\ref{section:summary}.

\section{Background}\label{section:background}

In this section, we present some necessary background and related work that provide the foundation for our proposed work in the next section.

\subsection{Spectrally Shaped Binary Sequence Design}\label{section:SSBS}

In previous work, we developed spectrally shaped sequences to provide a passband and notch for the pre-determined message and interferer bands, respectively~\cite{Mo2015Design-of-Spect, Mo:2018a}. We denote by $\vf _m = \left[1, e ^{j 2 \pi \left(m - 1\right) / N}, \dots, e ^{j 2 \pi \left(N - 1\right) \left(m - 1\right) / N} \right] ^T / \sqrt{N} $ ($m = 1, 2, \dots, N$) the $N$-dimensional complex-valued vector to represent the basis element of the discrete Fourier transform. We also denote by $\mf _\cP$ and $\mf _\cS$ the matrices whose columns provide orthonormal bases for the message band $\cP \subseteq \{1, 2, \dots, N\}$ and interferer band $\cS \subseteq \{1, 2, \dots, N\}$, respectively. For example, we can set $\mf _\cP$ and $\mf _\cS$ to have columns of three vectors $\vf _m$ for different disjoint intervals of the indices $m$. Then $\left \| \mf _\cP \vs \right\| _2 ^2$ and $\left \| \mf _\cS \vs \right \| _2 ^2$ can be used to measure the sequence power in the message and interferer bands, respectively. In order to obtain the spectrally shaped sequence, the sequence power in the message band should be as large as possible while keeping the sequence power in the interferer low. Our approach for designing an $N$-point binary sequence can be written as
\begin{align}\label{eq:single_general}
\hs = \arg \max _{\vs \in \bR^N} \quad
    & \left\| \mf ^H _\cP \vs \right\| _2 ^2 \notag \\
    \st \quad
    & \left\| \mf ^H _\cS \vs \right\| _2 ^2 \leq \alpha, \notag \\
    & s _k ^2 = 1, \quad k = 1, 2, \dots, N,
\end{align}
for some interferer tolerance $\alpha > 0$, where $s _k$ denotes the $k^{\mathrm{th}}$ entry of $\vs$. Such an integer optimization problem is a QCQP since both the objective function and constraints have a quadratic form with respect to $\vs$. This problem is also NP-hard. Though it is possible to use an exhaustive binary search when the sequence length is very small, it is too inefficient and even impossible to use the exhaustive method when the sequence length is relatively large.

We can solve the QCQP (\ref{eq:single_general}) using a semidefinite program (SDP) relaxation and randomized projection. By lifting $\vs$ to $\mt = \vs \vs ^T$, the SDP equivalent for the QCQP (\ref{eq:single_general}) can be obtained by noting that $\left\| \mf _\cP ^H \vs \right\| _2 ^2 = \trace{ \mf _\cP \mf _\cP ^H \vs \vs ^T} = \trace{\mf _\cP \mf _\cP ^ H \mt}$ and $\left\| \mf _\cS ^ H \vs \right\| _2 ^2 = \trace{ \mf _\cS \mf _\cS ^H \vs \vs ^T} = \trace{\mf _\cS \mf _\cS ^ H \mt}$, where $\trace{\cdot}$ represents the trace operator on a matrix. The binary constrain can be written as $T _{k, k} = 1$ and $\rank{\mt} = 1$, where $T _{k, k}$ denotes the $k ^\mathrm{th}$ diagonal entry of $\mt$ and $\rank{\cdot}$ represents the rank operator for a matrix. Any sequence $\vs$ that is feasible to the binary constrain can build a matrix $\mt = \vs \vs ^T$ that is feasible to the rank constraints. Furthermore, any feasible matrix $\mt$ with $\rank{\mt} = 1$ can be factorized as the product of a binary vector $\vs$ and its transpose by Cholesky decomposition or eigendecomposition. Dropping the rank constraint reduces the QCQP equivalent to a convex SDP
\begin{align}\label{eq:single_SDP}
    \hT = \arg \max _{\mt \in \mathbb{S} ^N} \quad
    & \trace{\mf _\cP \mf _\cP ^H \mt} \notag \\
    \st \quad
    & \trace{\mf _\cS \mf _\cS ^H \mt} \leq \alpha, \notag \\
    & T _{k, k} = 1, \quad k = 1, 2, \ldots, N,
\end{align}

After solving the SDP relaxation (\ref{eq:single_SDP}) and obtaining the optimal matrix $\hT$, unless $\hT$ has rank one, an approach is needed to obtain vectors $\vw$ whose Gram matrix $\vw \vw ^T$ are approximations to $\hT$. We rely on the method proposed by Goemans and Williamson to obtain approximations by randomized projection~\cite{goemans1995improved}. Assume that the \emph{eigenvalue decomposition} (EVD) of $\mt$ is $\mt = \mathbf{U} \boldsymbol{\Sigma} \mathbf{U} ^T$, where the unitary matrix $\mathbf{U}$ contains eigenvectors of $\mt$ in its columns and the diagonal matrix $\boldsymbol{\Sigma}$ contains eigenvalues in its diagonal entries. For each approximation, a vector $\vv$ is generated by drawing all entries independently and identically from the standard Gaussian distribution, i.e., $\vv \sim \mathcal{N} \left( \mathbf{0}, \mi \right)$, where $\mi$ represents the identity matrix. Then an approximation is the projection of $\hT$ onto the space spanned by the random vector $\vw = \mathbf{U} \boldsymbol{\Sigma} ^{1/2} \vv$, where $\boldsymbol{\Sigma} ^{1/2}$ is the element-wise square root of $\boldsymbol{\Sigma}$.

However, the approximations $\vw$ maybe not feasible to either the interferer or binary constraints in (\ref{eq:single_general}). Therefore, we perform binary quantization on the approximations, which forces the resulting sequence to be binary. In other words, a candidate sequence is obtained by $\ts = \sign{\vw}$, where $\sign{\cdot}$ represents the element-wise sign operator on a vector. We have shown that a candidate sequence fails to satisfy the interferer constraint with small probability~\cite{Mo:2018a}:
\begin{align}
    \prob{\left\| \mf _\cS ^H \ts \right\| _2 ^2 \ge \alpha} \le \exp \left( -C \frac{\alpha
    ^2}{\left| \cS \right|} \right),
\end{align}
where $\left| \cS \right|$ is the size of the interferer band $\cS$ and $C$ is a constant. The candidate sequence $\ts$ is an approximation to the optimal sequence $\hs$. When defining the approximation ratio as $r = \left\| \mf _\cP ^H \ts \right\| _2 ^2 / \trace{\mf _\cP \mf _\cP ^H \hT}$ to measure the performance of the objective value of $\ts$ with respect to that of $\hT$, the ratio also measures the performance with respect to $\ts$ since $\left\| \mf _\cP ^H \ts \right\| _2 ^2 / \left\| \mf _\cP ^H \hs \right\| _2 ^2 \ge r$ due to the fact that $\hs \hs ^T$ is a feasible solution to the SDP relaxation (\ref{eq:single_SDP}) and thus $\left\| \mf _\cP ^H \hs \right\| _2 ^2 \le \trace{\mf _\cP \mf _\cP ^H \hT}$. We have numerically shown that the approximation ratio for sequence design is no less than $r \ge \pi / 2 - 1$.

The proposed sequence design via SDP relaxation and randomized projection is named as \emph{randomized SDP relaxation} (RSDPR) and is detailed in Algorithm~\ref{alg:single_design}. The algorithm repeats the random projection $L$ times to provide a set of candidate sequences and finally outputs the sequence with maximum power in the message band among all that meet the requested upper bound for the sequence power in the interferer band.

\begin{algorithm}[t]
\renewcommand{\algorithmicrequire}{\textbf{Input:}}
\renewcommand{\algorithmicensure}{\textbf{Output:}}
\caption{Randomized SDP Relaxation}\label{alg:single_design}
\begin{algorithmic}[1]
    \REQUIRE{bases $\mf _\cP$ and $\mf _\cS$ for message and interferer bands, interferer tolerance
    $\alpha$, random search size $L$}
    \ENSURE{binary sequence $\hs$}
    \STATE{obtain optimal solution $\hT$ to SDP relaxation (\ref{eq:single_SDP})}
    \STATE{compute EVD for $\hT = \mathbf{U} \boldsymbol{\Lambda} \mathbf{U} ^T$}
    \FOR{$\ell = 1, 2, \ldots, L$}
    \STATE{generate random vector $\mathbf{v} \sim \mathcal{N} \left( \mathbf{0}, \mathbf{I} \right)$}
    \STATE{obtain approximation by projecting $\vw _\ell = \mathbf{U} \boldsymbol{\Lambda} ^{1/2} \vv$}
    \STATE{obtain candidate by quantization $\tilde{\vs} _\ell = \sign{\mathbf{w} _\ell}$}
    \ENDFOR{}
    \STATE{select best binary sequence}
\begin{align*}
    \hs = \arg \max _{\tilde{\vs} _\ell : 1 \le \ell \le L} \left\{ \left\| \mf _\cP ^H \ts _\ell
    \right\| _2 ^2 : \left\| \mf _\cS ^H \ts _\ell \right\| _2 ^2 \le \alpha \right\}
\end{align*}
\end{algorithmic}
\end{algorithm}

\subsection{Slepian Basis}\label{section:slepian}

The Slepian basis is often used to deal with spectrum leakage when narrowband signals contain arbitrary frequency components. We define an $N$-dimensional complex exponential vector as
\begin{align}
    \cF \left( f \right) = \frac{ 1 }{ \sqrt{ N } } {\left[1, e ^{ j 2 \pi f }, \dots, e ^{ j 2 \pi ( N - 1 ) f } \right] } ^T
\end{align}
where $f \in \cM = [0, 1]$ is the corresponding normalized frequency. The elements of the Fourier basis $\vf _m = \cF \left( f _m \right)$ ($m = 1, 2, \dots, N$) sample the normalized frequency range $\mathcal{M}$ uniformly with the sampled frequencies $f _m = (m - 1) / N \in \cM$, which we also refer to as the on-grid frequencies, while all other frequencies $f \in \cM$ are referred to as the off-grid frequencies. The fact that the complex exponentials at these off-grid frequencies cannot be sparsely represented using the Fourier basis elements is a behavior known as spectral leakage in the literature; such spectral leakage often negatively affects the performance of random demodulators with real-world signals.

In contrast, it is well-known that any bandlimited signal must be infinite in the time domain and no signal with finite length in the time domain can be bandlimited. In~\cite{Slepian1976On-bandwidth, Slepian1983Some-Comments-o}, Slepian provided remarkable work to approximate a bandlimited signal with \emph{discrete prolate spheroidal sequences} (DPSSs) in discrete time. However, DPSSs are of infinite length. Recently, the fast Slepian Transform was proposed for fast and efficient computation of approximated projections onto the leading Slepian basis elements~\cite{Karnik2016The-fast-Slepia, Karnik2016Fast-computatio}, which makes the Slepian basis a fully competitive alternative to the Fourier basis.

Given the length $N$ and the half bandwidth $W \in (0, 0.5)$ in normalized frequency, the DPSSs are a collection of $N$ discrete-time infinite-length signals that are strictly bandlimited to the normalized frequency range $[-W, W]$ but highly concentrated to their first $N$ entries in the time domain. The DPSSs are defined to be the eigenvectors of a procedure that suppresses all except the first $N$ entries of an infinite-length signal to zero and then filters out the frequency components of the length-$N$ signal outside the frequency range $[-W, W]$. The Slepian basis elements $\vg _1, \vg _2, \dots, \vg _N$ are defined as the time-limited DPSSs, which keeps only the first $N$ entries of those infinite-length DPSSs. The Slepian Basis forms an orthonormal basis for an optimal subspace approximation to the class of bandlimited signals.

\begin{figure}[t]
\centering
\includegraphics[width = 0.45\textwidth]{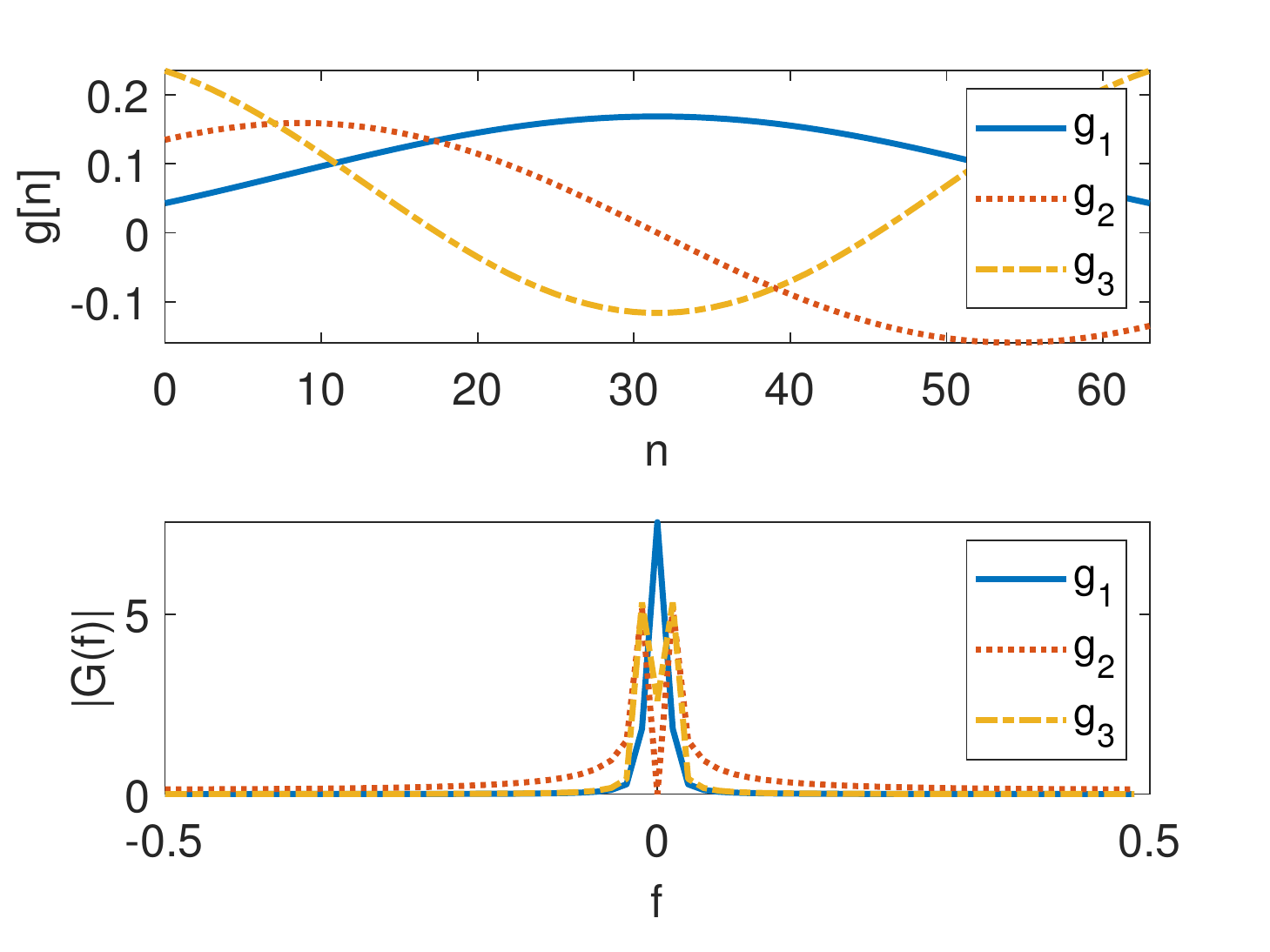}
\caption{(Top) First three elements of the Slepian basis $\vg_1[n], \vg_2[n], \vg_3[n]$ and (Bottom) their corresponding magnitude responses when $N = 64$ and $W = 1/N$. Though the elements of the Slepian basis have  finite length, their spectra are highly concentrated on the interval $[-W, W]$.}\label{fig:back_slepian}
\end{figure}

It has been shown that the first $2NW$ elements of the Slepian basis are usually sufficient to express the $N$-length samples of any signal bandlimited to the frequency range $[-W, W]$~\cite{Karnik2016The-fast-Slepia, Karnik2016Fast-computatio}, which is illustrated in Figure~\ref{fig:back_slepian}. By modulating the Slepian basis for baseband with an element of the Fourier basis, we can obtain a subspace approximation of signals restricted to any frequency subset of $[0, 1]$. For example, the modulated Slepian basis $\cF (f) \circ \vg _1, \cF (f) \circ \vg _2, \dots, \cF (f) \circ \vg _N$ can be used to represent a signal bandlimited to frequency range $[f - W, f + W]$, where $\circ$ denotes the element-wise product of two vectors. We can also concatenate such bases for different frequency ranges to obtain a frame that can successfully approximate multi-band signals of interest.

\begin{figure}[t]
\centering
\includegraphics[width = 0.45\textwidth]{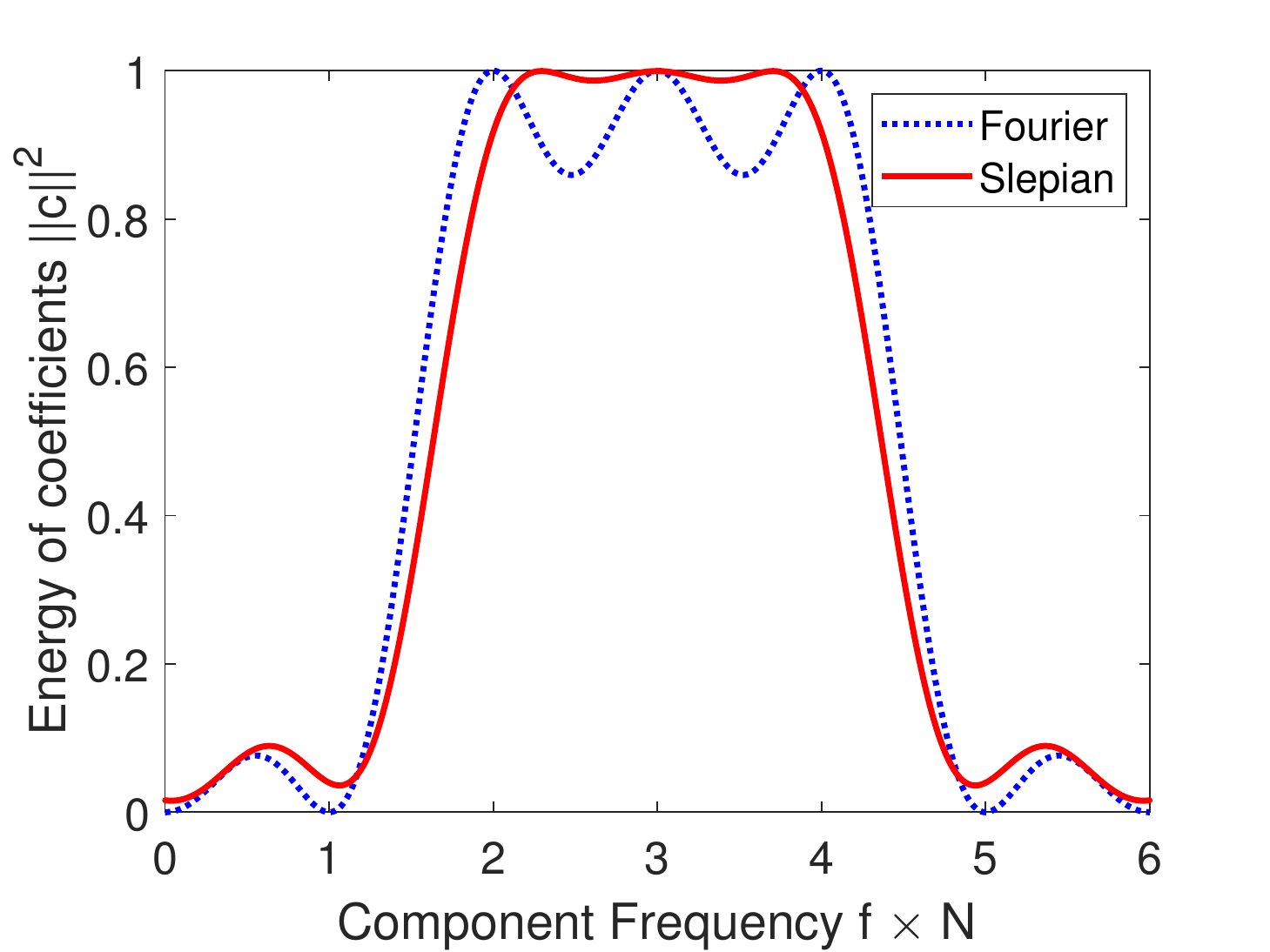}
\caption{Energy of the coefficient vector $\vc$ for a complex exponential signal as a function of its frequency under both choices of basis. While the Slepian basis successfully captures the energy of the signals, for most frequencies of interest, it also captures more energy for other on-grid frequencies, while the Fourier basis performs better rejection of the latter.}\label{fig:back_spec}
\end{figure}

Both the Slepian and Fourier bases can be used to form an orthonormal basis for a subspace approximation to the set of signals bandlimited to $\left[ -W, W \right]$. The most significant difference between the Fourier and Slepian representations appears for signals containing off-grid frequencies. For example, when $W = 1/N$, a complex exponential $\vx = \cF(f)$ for $f \in [f _{m - 1}, f _{m + 1}]$ can be approximated as a linear combination of three elements of either the Fourier basis $\mm = [\vf _{m - 1}, \vf _m, \vf _{m + 1}]$ or the three elements of the Slepian basis $\mm = \left[ \vf _{m} \circ \vg _1, \vf _{m} \circ \vg _2, \vf _{m} \circ \vg _3 \right]$ with coefficients $\vc = \mm ^H \vx$. Figure~\ref{fig:back_spec} shows the energy of the coefficient vector $\vc$ for a complex exponential signal as a function of its frequency under both choices of basis, with $m = 4$. Although the Fourier basis compacts the signal energy to a single coefficient when the frequency is on-grid (i.e., $f \in \{f _{m - 1}, f _m, f _{m + 1}\}$), some energy is leaked to other coefficients when the frequency is off-grid. In contrast, the top three coefficients of the signal in the Slepian basis capture almost all of the energy of the signal at all values of the frequency within the band of interest. Nonetheless, the Fourier basis has better rejection than the Slepian basis for signals with on-grid frequencies outside the bandwidth of interest, which also affects its suitability to model signals restricted to a bandwidth within our design approach.

\section{Multi-Branch Sequence Set Design}

In this section, we present the details of our proposed approach for multi-branch sequence set design. Our approach is based on our prior work for single sequence design, which is used iteratively to obtain sequences for different branches and introduce an approximate orthogonality constraint. We also analyze the relationship between the condition number and the orthogonality tolerance value. Our analysis shows that it is impossible to obtain a set of mutually orthogonal sequences of size equal to the sequence length in order to provide stable recovery, and so it is necessary to introduce oversampling in the sequence design.

\subsection{Sequence design}\label{section:design}

We seek a set of binary sequences $\vs _1, \vs _2, \dots, \vs _N \in \{-1, 1\} ^N$ that are used to modulate the received signals in a multi-branch modulation architecture. In the modulation, the interferer band should be suppressed as much as possible; after the modulation, the message band should be recovered from the multiple modulations of the signal. To mitigate the interferer, the binary sequence modulation should work as a band-stop filter that provides a notch at the interferer band. Therefore, the sequences shall have small power in the interferer band. In order to obtain a stable reconstruction of the message, the modulation system involving the sequence set should be well-conditioned to prevent large distortion in the output due to noise. As we will show in the next subsection, the requirement for the sequences to be as close to mutually orthogonal as possible provides a guarantee that the sequence set has small condition number.

We denote by $\mf _\cS$ and $\mf _\cP$ the collection of basis elements for the interferer and message bands, respectively. We also assume that $\cS$ and $\cP$ are disjoint, i.e., there is no overlap between the message and interferer bands. The power of a sequence $\vs _k$ ($k = 1, 2, \dots, N$) in the interferer band can be measured by $\| \mf _\cS ^H \vs _k \|_2 ^2$. The orthogonality between a pair of sequences $\vs _i$ and $\vs _j$ can be measured by the normalized inner product $\left< \vs _i, \vs _j \right> / \left( \left\| \vs _i \right\| _2 \left\| \vs _j \right\| _2 \right) = \vs _i ^T \vs _j / N$ due to every binary sequence satisfying $\left\| \vs _i \right\|_2 = \sqrt{N}$.

The sequence design problem for each channel is then to find a binary sequence $\vs$ such that the corresponding sequence power in the interferer band is minimized while the inner product between the sequence and each previously designed sequence for other channels is sufficiently small. Thus an approach for designing the sequence for $k ^{\mathrm{th}}$ channel ($k = 1, 2, \dots, N$) can be written as the QCQP
\begin{align}\label{eq:plain}
    \hs _k = \arg \min _{\vs \in \bR ^N} \quad
    & \left\| \mf _\cS ^H \vs \right\| _2 ^2 \notag \\ 
    \st \quad
    & \left| \hs _i ^T \vs \right| ^2 \le \alpha N, i = 1, 2, \dots, k - 1, \notag \\
    & s _n ^2 = 1, n = 1, \dots, N,
\end{align}
where $\alpha$ is the orthogonality tolerance, and $\hs _i$ ($i = 1, \dots, k - 1$) is the obtained sequence for the $i ^{\mathrm{th}}$ channel.

It is easy to verify that (\ref{eq:plain}) is a QCQP due to the fact that $\left| \hs _i ^T \vs _k \right| ^2 = \vs _k ^T \hs _i \hs _i ^T \vs _k = \trace{\hs _i \hs _i ^T \vs _k \vs _k ^T}$. Following the framework prescribed in Section~\ref{section:SSBS}, the sequence design can be approximately solved by a SDP relaxation and randomized projection. Before we present the details about solving (\ref{eq:plain}), some discussion about the condition number for the equivalent modulation operator matrix of the obtained sequence set is necessary.

\subsection{Condition Number}\label{section:condition}

From~\cite{laska2007theory, Mishali2009Blind-Multiband, tropp2010beyond}, the modulation of the input signal $\vx$ can be simplified as $\vy = \ms \vx$ when noise and nonlinearity are ignored, where $\ms = {[\hs _1, \dots, \hs _N]} ^T$ represents the sequence set and $\vy$ is a vector containing the modulation samples after integration for the different channels. When $\ms$ is invertible, the original sequence can be recovered by $\hat{\vx} = \ms ^{-1} \vy$.

When there is noise or distortion $\ve$ added to the receiver samples $\vy$ before recovery, the error in the output will be $\ms ^{-1} \ve$. Thus, the recovery performance can be measured by the proportion of the signal-to-noise ratios (SNRs) before and after recovery:
\begin{align}\label{eq:con}
    \frac{ \left. \left\| \vy \right\| _2 \right/ \left\| \ve \right\| _2 }{ \left. \left\| \ms ^{-1} \vy \right\| _2 \right/ \left\| \ms ^{-1} \ve \right\|_2 }
    &=\frac{ \left\| \ms ^{-1} \ve \right\|_2 }{ \left\| \ve \right\|_2 } \frac{ \left\| \vy \right\|_2 }{ \left\| \ms ^{-1} \vy \right\| _2 } \notag \\
    &=\frac{ \left\| \ms ^{-1} \ve \right\|_2 }{ \left\| \ve \right\|_2 } \frac{ \left\| \ms \vx \right\|_2 }{ \left\| \vx \right\| _2 }
\end{align}
Smaller values for this ratio indicate better recovery performance. The condition number of $\ms$ can be similarly defined as the maximum possible value of the ratio product
\begin{align} 
    \kappa = \max _{ \ve \ne \mathbf{0} } \frac{ \left\| \ms ^{-1} \ve \right\| _2 }{ \left\| \ve \right\| _2 } \max _{ \vx \ne \mathbf{0} } \frac{ \left\| \ms \vx \right\|_2 }{ \left\| \vx \right\| _2 },
\end{align}
which shows the maximum possible error occurring in the recovery. The larger that the condition number is, the worse that the recovery can potentially be.

We denote the singular value decomposition of $\ms = \mathbf{U} \boldsymbol{\Sigma} \mv^T$, where $\mathbf{U} = \left[ \vu _1, \dots, \vu _N \right]$ and $\mv = \left[ \vv _1, \dots, \vv _N \right]$ are both unitary matrices and $\boldsymbol{\Sigma}$ is a diagonal matrix whose diagonal entries are the singular values $\sigma _1 \ge \dots \ge \sigma _N > 0$. Then
\begin{align}
    \max _{\vx \ne \mathbf{0}} \frac{\left\| \ms \vx \right\|_2}{\left\| \vx \right\| _2} = \sigma _1,
\end{align}
where the equality is satisfied if and only if $\vv _1 ^T \vx = 1$ and $\vv _i ^T \vx = 0$ for $i = 2, \dots, N$, i.e., $\vx = \vv _1$. 

Since both $\mathbf{U}$ and $\mv$ are unitary, $\ms ^{-1} = {\left( \mathbf{U} \boldsymbol{\Sigma} \mv \right)} ^{-1} = \mv ^T \boldsymbol{\Sigma} ^{-1} \mathbf{U}$, where $\boldsymbol{\Sigma} ^{-1}$ is a diagonal matrix whose diagonal entries are $1 / \sigma _N \ge \dots \ge 1 / \sigma _1 > 0$. Similarly, 
\begin{align}
    \max _{ \ve \ne \mathbf{0} } \frac{ \left\| \ms ^{-1} \ve \right\| _2 }{ \left\| \ve \right\| _2 } = \frac{ 1 }{ \sigma _N }.
\end{align}
Thus the condition number is also equal to $\kappa = \frac{\sigma _1} {\sigma _N}$. In other words, the condition number is also defined as the ratio between the largest and smallest singular values. When the condition number is infinity, $\sigma _N = 0$ and $\ms$ is not invertible. When $\sigma _N$ is close to zero rather than strictly equal to zero, the condition number is extremely large. The error after recovery is very large even if the error before recovery is very small; such recovery is not stable. The closer that the minimum singular value is to zero, the worse that the recovery is. To guarantee a stable recovery, it is required that the minimum singular value $\ms$ is far away from zero.

We can leverage the relationship between the minimum singular value of $\ms$ or the minimum eigenvalue of the Gram matrix $\mq = \ms \ms ^T$, whose diagonal entries $Q _{i, i} = \hs _i^T \hs _i = N$ due to the binary entries of $\hs _i$, and all off-diagonal entries $Q _{i, j} = \hs _i ^T \hs _j$ are upper-bounded in the sequence design. In~\cite{Gerschgorin:1931}, Gershgorin proved the following theorem to reveal the relationship between the eigenvalues and entries of a matrix.

\begin{theorem}[Gershgorin Circle Theorem]\label{th:gc}
For a square matrix $\mq \in \bC ^{N \times N}$, let $R _i = \sum _{j \ne i} |Q _{i, j}|$ ($i = 1, 2, \dots, N$) be the sum of the absolute values of the off-diagonal entries in the $i ^\mathrm{th}$ row, and $\mathcal{D}\left(Q _{i, i}, R _i\right) \subset \bC$ be a closed disc centered at $Q _{i, i}$ with radius $R _i$, which is called a Gershgorin disc. Then each eigenvalue of $\mq$ lies within at least one of the Gershgorin discs $\mathcal{D} \left(Q_{i, i}, R _i\right)$.
\end{theorem}

In other words, there exists at least one index $i = 1, \dots, N$ such that the minimum singular value satisfies $\left|\sigma _N ^2 - Q _{i, i}\right| \le \sum _{j \ne i} \left|Q _{i, j}\right|$. If there exists $i \in \left\{1, \dots, N\right\}$ such that $\sum _{j \ne i} \left|Q _{i, j}\right| \ge Q _{i, i} = N$, which implies that the corresponding Gershgorin discs of $\mq$ contains the origin, then the minimum singular value of $\mq$ could be arbitrarily close to zero, and so the condition number could be arbitrarily large. In order to have stable recovery, the value of the off-diagonal entries should be as small as possible, which indicates that sequences should be as close to mutually orthogonal as possible.

For a pair of binary sequences $\vs _i, \vs _j \in \left\{ -1, 1 \right\} ^N$, $\left< \vs _i, \vs _j \right> = \sum _{n = 1} ^N s _{i, n} s _{j, n}$, where $s _{i, n}$ is the $n ^\mathrm{th}$ entry of $\vs _i$. Since the product term $s _{i, n} s_{j, n}$ also takes a binary value, the pair of binary sequences are orthogonal, i.e., $\left< \vs _i, \vs _j \right> = 0$ if and only if $N/2$ pairs of entries of $\vs _i$ and $\vs _j$ at the same indices have the same values, and the other pairs of entries of $\vs _i$ have the opposite values. If there are more or less than $N / 2$ matching pairs of entries, then the numbers of positive and negative terms do not match.

When $N$ is odd, it is obviously impossible for a pair of binary sequences $\vs _i$ and $\vs _j$ to be be orthogonal; additionally, $\left| \left< \vs _i, \vs _j \right> \right| \ge 1$. Thus, the sum of absolute values of all off-diagonal entries $\sum _{ i\ne j } \left| Q _{i, j} \right| \ge N - 1$ for all rows. According to Theorem~\ref{th:gc}, the minimum eigenvalue may be as small as $1$ and the maximum eigenvalue may be as large as $2N-1$, making the condition number of $\mq$ as large as $2N-1$ even in this case where all sequences are as orthogonal to each other as possible. 

There are special cases of orthogonal binary sequence sets when $N$ is a power of $2$: a well-known example is the Walsh-Hadamard codes. These codes are constructed from the elementary matrix
\begin{align}
    \mh _2 = \begin{bmatrix} 1 & 1 \\ 1 & -1 \end{bmatrix}.
\end{align}
This so-called \emph{Hadamard matrix} contains the two codewords in the $2$-dimensional Walsh-Hadamard codes. Higher-dimensional Walsh-Hadamard codes can be constructed using the Hadamard matrix as follows:
\begin{align}
    \mh _n = \begin{bmatrix} \mh _{n / 2} & \mh _{n / 2} \\ \mh _{n / 2} & -\mh _{n / 2} \end{bmatrix}
\end{align}
The matrix $\mh_n$ contains $n$ orthogonal binary codewords of length $n$. Although the Walsh-Hadamard codes in multi-branch modulation provide a perfect recovery, their construction is binary only when $N$ is a power of $2$. Additionally, the Walsh-Hadamard codes are fixed and its interferer mitigation cannot be tailored to prior knowledge of the interferer band. In fact, it is straightforward to verify that the Hadamard codes for $N$-dimensional space contains the sequences ${ \left[ 1, 1, \dots, 1 \right] }^T$ and ${ \left[ 1, -1, \dots, -1\right ] } ^T$. Those two sequences have non-zero spectra only at normalized frequencies $0$ and $1/2$, respectively. Therefore, the Walsh-Hadamard codes mitigation performance suffers when the interferer bands contain those frequencies.

Although an analytical proof of the difficulty of the design of approximately mutually orthogonal binary sequences remains elusive, we will numerically explore the feasibility of the QCQP (\ref{eq:plain}) that aims to find $N$ binary sequences of length $N$ that are approximately mutually orthogonal.

\subsection{Oversampling}\label{section:oversampling}

The analysis above shows that it is hard to obtain $N$ binary sequences in $N$-dimensional space that is approximately orthogonal such that the Gershgorin discs are far away from the origin. Intuitively, it is easier to find $N$ binary sequences that are approximately mutually orthogonal in a higher-dimensional space. Thus, we assume that each sequence has length $RN$, which can be used to modulate signals oversampled by a factor of $R$.

As described in Section~\ref{section:SSBS}, we obtain candidate sequences $\vs _\ell$ from the solution of a SDP relaxation $\hT$ via randomized projection and binary quantization. In summary, we generate random vectors $\vw_\ell$ as independent samples from a multivariate Gaussian distribution and the candidate $binary$ sequences are obtained via the quantization $\vs _\ell = \sign{\vw _\ell}$. While the presence of $\hT$ in the design of the sequences introduces correlations, we study the simpler case in which the sequence entries are independent, i.e., when $\hT = \mi$ and $\vw _\ell \sim \cN \left(\mathbf{0}, \mi\right)$. In this case, the entries of $s _{\ell, n} = \sign{w _{\ell, n}}$ ($n = 1, 2, \dots, RN$) are random variables drawn independently and identically from a Rademacher distribution, i.e., $\prob{s _{\ell, n} = 1} = \prob{s _{\ell, n} = -1} = 1 / 2$, where $\prob{\cdot}$ returns the probability of an event. Additionally, the pairwise entry products involved in the computation of the inner products of $\vs _i$ and $\vs _j$ also follow a Rademacher distribution, i.e., $\prob{s _{i, n} s _{j, n} = 1} = \prob{s _{i, n} s _{j, n} = -1} = 1 / 2$ for any $i \ne j$. The following theorem shows an upper for a so-called Rademacher sum.

\begin{theorem}[\cite{Montgomery-Smith:1990}]\label{thm:SSBS_prob}
Assume that $s _1, s _2, \dots, s _N$ is a sequence of random variables following a Rademacher distribution and $x _1, x _2, \dots, x _N$ is a set of real numbers. Then $\prob{\sum _{n = 1} ^N x _n s _n \ge t \sqrt{\sum _{n = 1} ^N x _n ^2}} \le e ^{-t ^2 / 2}$ for any $t \ge 0$.
\end{theorem}

To apply Theorem~\ref{thm:SSBS_prob} to the study of the inner product $\vs _i$ and $\vs _j$, we set $t = \alpha \sqrt{RN}$ and $x _n = 1$ for $n = 1, \dots, RN$ to obtain the probabilities 
\begin{align*}
\prob{\sum _{n = 1} ^{RN} s _{i, n} s _{j, n} \ge \alpha RN} &\le e ^{-\alpha ^2 RN / 2},\\
\prob{\sum _{n = 1} ^{RN} s _{i, n} s _{j, n} \le -\alpha RN} &\le e ^{-\alpha ^2 RN / 2},
\end{align*} 
where the latter statement is obtained by symmetry. Thus,
\begin{align}\label{eq:SSBS_feasibility}
\prob{\left| \vs _i ^T \vs _j \right| \ge \alpha RN} 
&= \prob{\left| \sum _{n = 1} ^{RN} s _{i, n} s _{j, n} \right| \ge \alpha RN} \notag \\ 
&\le 2 e ^{-\alpha ^2 RN / 2}.
\end{align}

Equation (\ref{eq:SSBS_feasibility}) shows that the probability that the pair of binary sequences $\vs _i$ and $\vs _j$ is not approximately orthogonal decreases exponentially with the oversampling rate $R$. With a higher oversampling rate, we are more likely to find approximately orthogonal sequences. It is difficult to derive a similar result to (\ref{eq:SSBS_feasibility}) when the sequences are drawn according to the solution of the SDP relaxation since the resulting sequences failed to be independent. Nonetheless, the numerical results in the sequel confirm the conclusion that increasing the oversampling helps to obtain sequences with better interferer performance.

However, the signals cannot be recovered when the oversampled sequences are used to modulate the signals, given that the columns of the resulting modulation matrix operator $\ms \in \left\{-1, 1\right\} ^{N \times RN}$ are not linearly independent. In order to address the ambiguous in restruction, we redefine the complex exponential vector for the normalized frequency $f \in \cM$ in the oversampled space as
\begin{align}
\cF \left( f \right) = \frac{ 1 }{ \sqrt{RN} } \left[ 1, e ^{ j 2 \pi f }, \dots, e ^{ j 2 \pi (RN - 1) f } \right].
\end{align}
The Fourier basis elements $\vf _m$ ($m = 1, 2, \dots, RN$) corresponding to the complex exponential vectors with the on-grid frequencies $f _m = \left(m - 1\right) / RN \in \cM$ sample the normalized frequency range more finely than those in the original space. Due to the oversampling, the original frequency range $\left[ 0, 1 \right]$ for $N$-dimensional signals is mapped to the frequency range $\left[ 0, 1/R \right]$ for $RN$-dimensional signals. Thus, while discussing the oversampled signal representations, we focus on signals that lie in the normalized frequency range $\left[ 0, 1/R \right]$, which contains the on-grid frequencies $f _1, f _2, \dots, f _N$.

Note that the signal $\vx$ can be expressed as the linear combination of the basis elements for the message band and interferer bands, i.e., $\vx = \mf _\cP \vc _\cP + \mf _\cS \vc _\cS$, where $\vc _\cP$ and $\vc _\cS$ are the corresponding basis coefficients. Then the observations $\vy = \ms \vx = \ms \mf _\cP \vc _\cP + \ms \mf _\cS \vc _\cS$. When the message band does not cover the whole spectrum, i.e., $\left| \cP \right| < N$, $\ms \mf _\cP$ has linearly independent columns, and so it is possible to recover the coefficients via the pseudoinverse $\hat{ \vc } _\cP = { \left( \ms \mf _\cP \right) } ^{ \dagger } \vy = { \left( {\left(\ms \mf _\cP \right) } ^T \left( \ms \mf _\cP \right) \right) } ^{-1} { \left( \ms \mf _\cP \right) } ^T \vy$.

In the observations, the error with respect to the message, denoted by $\ve ' = \ve + \ms \mf _\cS \vc _\cS$, consists of both an additional error $\ve$ from noise and nonlinearity and the interferer. Following an analysis similar to (\ref{eq:con}), the recovery performance for the message coefficients can be measured by the ratio of the input and output SNRs

{\small \begin{align*}
\frac{\left. \left\| \ms \mf _\cP \vc _\cP \right\| _2 \right/ \left\| \ve' \right\| _2 }{ \left. \left\| { \left(\ms \mf _\cP \right) } ^{\dagger} \left( \vy - \ve' \right) \right\| _2 \right/ \left\| { \left( \ms \mf _\cP \right) } ^{\dagger} \ve '\right\| _2 }
=\frac{ \left\| { \left( \ms \mf _\cP \right) } ^{\dagger} \ve' \right\| _2 }{ \left\| \ve' \right\| _2 } \frac{ \left\| \ms \mf _\cP \vc _\cP \right\| _2 }{ \left\| \vc _\cP \right\| _2}.
\end{align*}}
Thus, the recovery performance can be measured by the condition number of $\ms \mf _\cP$, i.e., the ratio between the maximum and minimum nonzero singular value of $\ms \mf _\cP$. This indicates that the sequence projections onto the message band are required to be as close to being orthogonal to each other as possible.

\subsection{Design Algorithm}\label{section:algorithm}

Based on the analysis in the previous subsection, the sequence design for the $k ^\mathrm{th}$ channel finds a binary sequence $\vs _k$ such that the corresponding sequence power in the interferer band is minimized while the inner products between the projections of each pair of sequences on the message subspace are sufficient small. Thus, it can be written as the QCQP
\begin{align}\label{eq:design}
\hs _k = \arg \min _{\vs \in \bR ^N} \quad
& \left\| \mf _\cS ^H \vs \right\| _2 ^2 \notag \\
\st \quad
& \left| \left<\mf _\cP ^H \hs _i , \mf _\cP ^H \vs \right> \right| ^2 \le \alpha RN, i = 1, 2, \dots, k - 1, \notag \\
& {s [n]} ^2 = 1, n = 1, \dots, RN.
\end{align}
The SDP relaxation for (\ref{eq:design}) can be obtained by noting that $\left\| \mf _\cS ^H \vs \right\| _2 ^2 = \trace{\mf _\cS \mf _\cS ^H \vs \vs ^T}$ and $ \left| \left<\mf _\cP ^H \hs _i, \mf _\cP ^H \vs \right> \right| ^2 = \trace{\mf _\cP \mf _\cP ^H \hs _i \hs _i ^T \mf _\cP \mf _\cP ^H \vs \vs ^T}$. By lifting $\vs$ to $\mt = \vs \vs ^ T$, the SDP relaxation is
\begin{align}\label{equation:SDP}
\hT _k = \arg \min _{\mt \in \bS ^{RN}} \quad
& \trace{\mf _\cS \mf _\cS^H \mt} \notag \\
\st \quad
& \trace{\mf _\cP \mf _\cP ^H \vs _i \vs _i ^T \mf _\cP \mf _\cP^H \mt} \le \alpha RN, \notag \\
& \quad \quad i = 1, 2, \dots, k - 1, \notag \\
& T_{n, n} = 1, \quad n = 1, 2, \dots, RN.
\end{align}

After obtaining each $\hT _k$, we use the randomized projection and binary quantization to extract a binary sequence $\hs _k$ for each channel. We called the resulting algorithm \emph{multi-branch randomized  SDP relaxation} (MB-RSDPR). As shown in Algorithm~\ref{alg:multi_design}, MB-RSDPR repeatedly generates a random vector $\vv$ to project $\hT _k$ from a matrix space to a vector space and obtain the approximation vector $\vw _\ell$. A candidate binary sequence $\tilde{\vs} _\ell$ is then obtained by applying binary quantization on the approximation vector $\tilde{\vs} _\ell = \sign{\vw _\ell}$. The MB-RSDPR repeats the random projection $L$ times to provide a set of candidate sequences and finally outputs the sequence $\hs _k$ that minimizes the objective function while satisfying the constraints in the QCQP (\ref{eq:plain}). All sequences are obtained iteratively by following the same process. Obviously, solving the SDP relaxation problem (\ref{equation:SDP}) dominates the complexity of MB-RSDPR. Based on the fact that the SDP relaxation problem (\ref{equation:SDP}) can be solved in $\cO \left( N ^{4.5} \right)$~\cite{Potra2000Interior-Point-, Helmberg2006An-Interior-Poi}, MB-RSDPR has a complexity of $\cO \left( N ^{5.5} \right)$ in the worst case.

It is important and necessary to generate a sufficiently large number of candidate sequences to meet the constraints and return the best sequence for each channel. The results in the following section show that the size of the proposed randomized search is far smaller than that of an exhaustive search.

\begin{algorithm}[t]
\renewcommand{\algorithmicrequire}{\textbf{Input:}}
\renewcommand{\algorithmicensure}{\textbf{Output:}}
\caption{\sl Multi-Branch Randomized SDP Relaxation\/}\label{alg:multi_design}
\begin{algorithmic}[1]
    \REQUIRE{interferer band basis $\mf _\cS$, message band basis $\mf _\cP$, coherence tolerance $\alpha$,
    sequence length $N$, oversampling factor $R$, number of randomized projections $L$}
    \ENSURE{sequences $\hs _1, \hs _2, \dots, \hs _N$}
    \FOR{$k = 1, 2, \dots, N$}
    \STATE{obtain optimal solution $\hT _k$ to SDP relaxation (\ref{equation:SDP})}
    \STATE{compute EVD for $\hT _k = \mathbf{U} \boldsymbol{\Lambda} \mathbf{U} ^T$}
    \FOR{$\ell = 1, 2, \ldots, L$}
    \STATE{generate random vector $\vv \sim \cN \left( \mathbf{0}, \mi \right)$}
    \STATE{obtain approximation by projecting $\vw _\ell = \mathbf{U} \boldsymbol{\Lambda} ^{1/2}
    \vv$}
    \STATE{obtain candidate by quantization $\ts _\ell = \sign{\vw _\ell}$}
    \ENDFOR{}
    \STATE{select best binary sequence}
\begin{align*}
    \hs _k = \arg \max _{\ts _\ell : 1 \le \ell \le L} \left\{ \left\| \mf _\cS ^H \ts _\ell
    \right\| _2 ^2 : 
    \tiny{
    \begin{aligned} 
    \left| \hs _i ^T \mf _\cP \mf _\cP ^H \ts _\ell \right| \le \alpha RN \\ 
    i = 1, 2, \dots, k - 1
    \end{aligned}
    }
     \right\}
\end{align*}
    \ENDFOR{}
\end{algorithmic}
\end{algorithm}

\subsection{Basis Choice}\label{section:BasisChoice}

In the sequence design described in (\ref{eq:design}), the message and interferer in the signals are always assumed to lie in the space spanned by the Fourier basis elements for the message and interferer bands, respectively. When there is no overlap between the message and interferer bands, all elements in the basis for the message band are orthogonal to those in the basis for the interferer band. If the signals contain components with only on-grid frequencies, the frequencies sampled by the Fourier basis, there is no distortion caused by the interferer in recovering the message when the system operates in the linear regime.

As mentioned in Section~\ref{section:slepian}, when a signal contains components with frequencies not belonging to the on-grid frequency set, the representation of the signal with the Fourier basis is not perfect and not all energy of the signal is captured by the basis projections. If the interferer in the signal has components with off-grid frequencies, then some distortion appears in the recovered message even if a nonlinearity is not present. Therefore, the sequence set designed using the Fourier basis has suboptimal performance when the interferer spectrum covers off-grid frequencies in its band.

The Slepian basis has been advertised as a suitable representation for any signal with frequencies that lie in a small range containing both on-grid and off-grid frequencies. However, the Slepian basis for a frequency range that is outside of baseband relies on modulation with a complex exponential component whose frequency is the center frequency of the range. When the complex exponential components for the message and interferer bands are not orthogonal, the interferer may cause some distortion during the message recovery.

\section{Numerical Experiments}\label{section:results}

We conduct several experiments to test the performance of the MB-RSDPR for the design of multi-branch binary sequences. In the following experiments, we fix both the sequence length and the number of channels to be $N = 15$. The oversampling rate varies in the range $R \in \left\{ 1, 2, \dots, 10 \right\}$. The half bandwidth of the interferer band is $W = 1/RN$ such that the band covers the frequency range $\cM _\cS = \left[ f _{c - 1}, f _{c + 1} \right] \subset \cM$, where the on-grid frequency $f _c = (c - 1) / RN$ ($c = 2, 3, \dots, N - 1$) is its center frequency and is randomly chosen in the following experiments. The message band covers the rest of the spectrum, i.e., $\cM _\cP = \left( f _1, f _{c - 1} \right) \cup \left( f _{c + 1}, f _{N} \right) \subset \cM$. We denote the indices of the on-grid frequencies that fall into the message and interferer band by $\cP = \left\{ i : f _i \in \cM _\cP \right\} = \left\{ 1, \dots, c - 2, c + 2, \dots, N \right\}$ and $\cS = \left\{ i : f _i \in \cM _\cP \right\} = \left\{ c - 1, c, c + 1 \right\}$, respectively.

We use two metrics for the performance of the obtained sequence set $\ms = {\left[ \hs _1, \dots, \hs _N \right]} ^T$. To measure the interferer mitigation, we use the normalized sequence power in the interferer band, i.e., $\left\| \ms \mf _\cS \right\| _F ^2 / RN$, where $RN$ represents the total power of the sequence set. To measure the recovery stability, we use the condition number of $\ms \mf _\cP$, the projection of the sequence set onto the message space. Additionally, when no sequence set candidate meeting the constraints of (\ref{eq:design}) is found, we set the interferer power and condition number to be infinity. Values of the interferer power above $0\mathrm{dB}$ and of the condition number above $20$ are shown as $0\mathrm{dB}$ and $20$, respectively, in the following figures.

\begin{figure}[t]
\centering
\includegraphics[width = 0.45\textwidth]{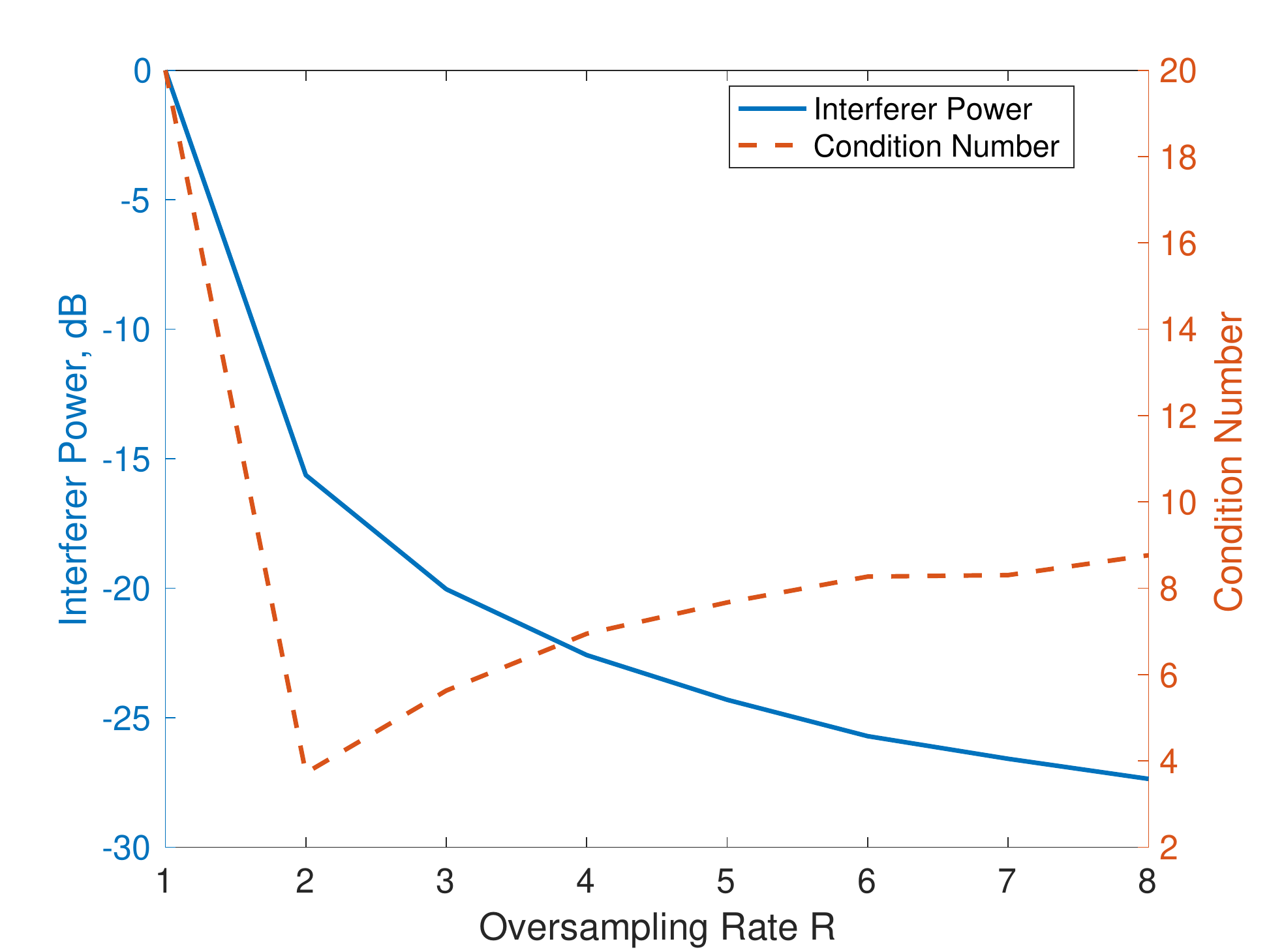}
\caption{Average interferer power and condition number as a function of the oversampling rate $R$. When there is no oversampling, both the interferer power and the condition number are outside the plotted range as no feasible sequence set was found. When oversampling is included, the interferer power decreases and the condition number increases as the oversampling rate increases.}\label{fig:over}
\end{figure}

In the first experiment, we fix the number of randomized projection $L = 10 ^5$ and coherence tolerance $\alpha = 0.4$. Figure~\ref{fig:over} shows the average interferer power and condition number of $100$ independently generated multi-branch sequence sets when the oversampling rate $R$ varies among $[1, 8]$. As we mentioned, when $R = 1$, i.e., there is no oversampling, it is hard to obtain $N$ binary sequences that are approximately orthogonal to each other. Thus, the condition number can be very large. When $R > 1$, i.e., oversampling is included in the sequence design, the interferer power decreases as the oversampling rate increases at the cost of an increasing condition number.

\begin{figure}[t]
\centering
\includegraphics[width = 0.45\textwidth]{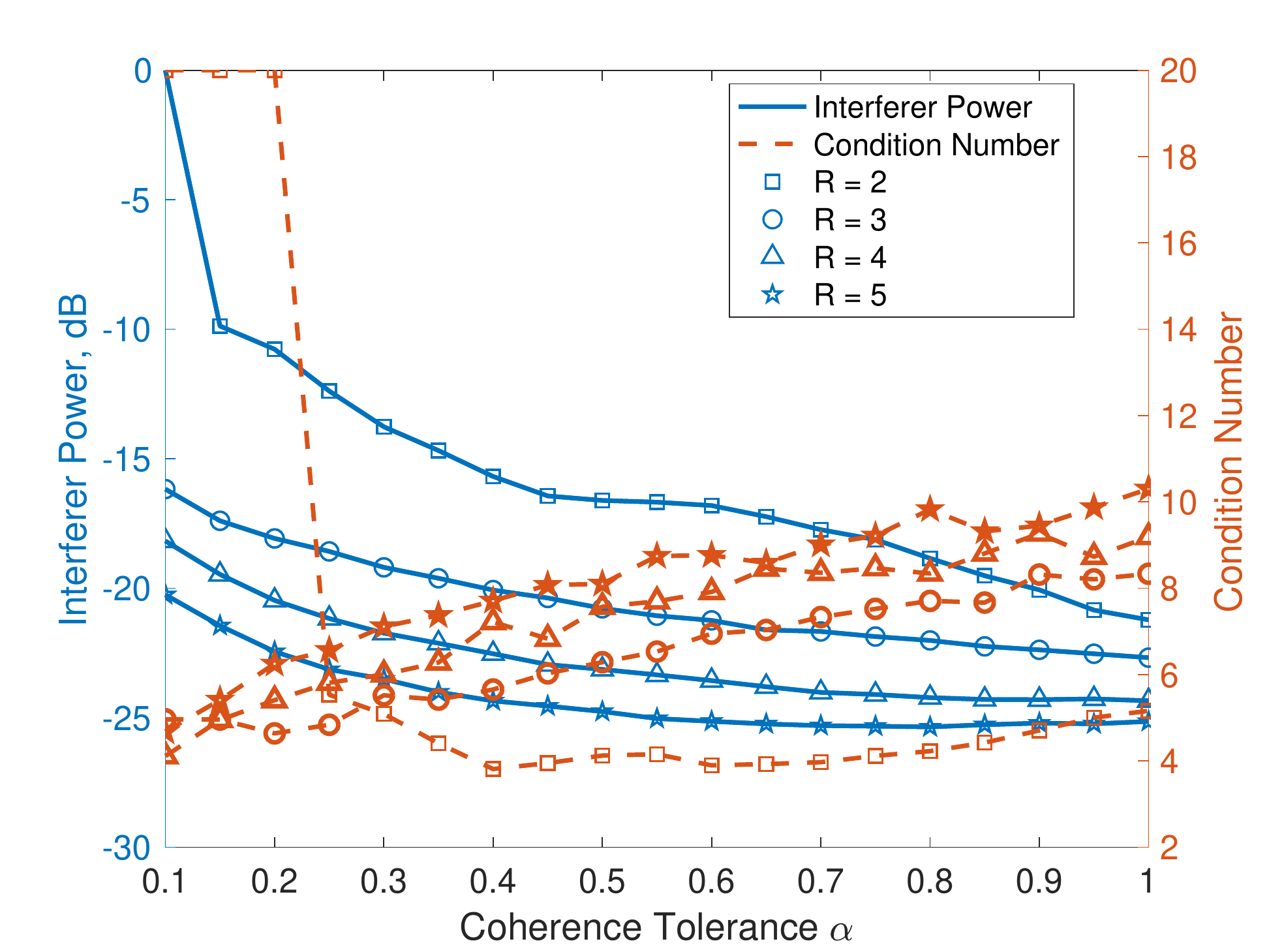}
\caption{Average interferer power and condition number as a function of the coherence tolerance $\alpha$. When R=2, setting the coherence tolerance to a small value makes it difficult to search for a feasible solution. For all oversampling rates, increasing coherence tolerance improves interferer mitigation performance but also increases the condition number.}\label{fig:tol}
\end{figure}

In the second experiment, we vary the coherence tolerance in the range of $\left[ 0.1, 1 \right]$. Figure~\ref{fig:tol} shows the corresponding average interferer power and condition number, and demonstrates that there is a trade-off between the interferer power and condition number: relaxing the coherence tolerance helps to obtain sequences with better interferer mitigation performance; however, the recovery performance decreases in turn. Note that when $R = 2$ and $\alpha$ is small, the condition number falls outside the plotted range. This again confirms the conclusion that it is hard to obtain a binary sequence set that is approximately mutually orthogonal in a low-dimensional space.

\begin{figure}[t]
\centering
\includegraphics[width = 0.45\textwidth]{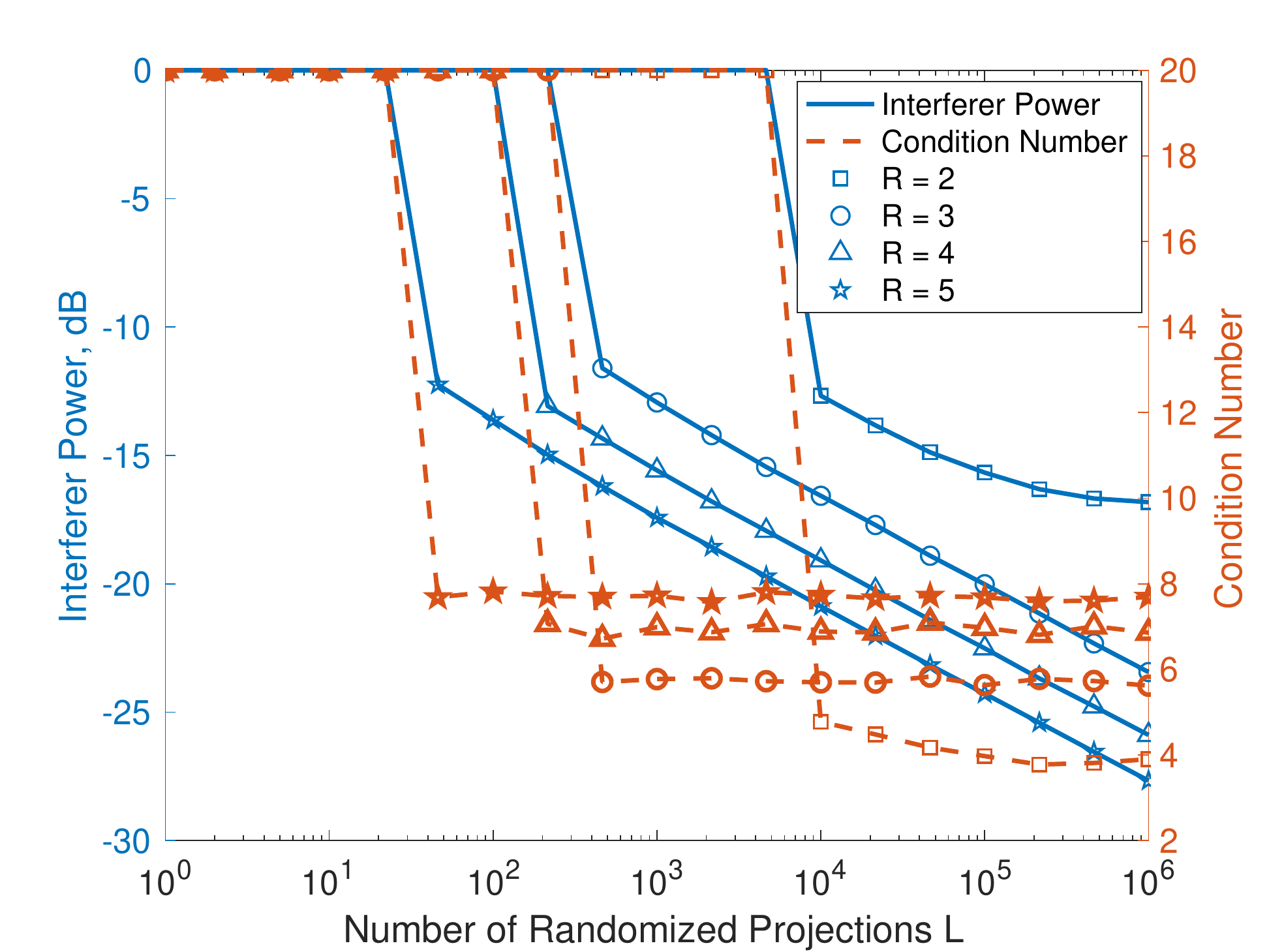}
\caption{Average interferer power and condition number as a function of the number of randomized projections $L$. The interferer power decreases as the number of randomized projections increases, while the condition number keep stable. The necessary number of randomized projections to obtain a feasible sequence set decreases as the oversampling rate increases.}\label{fig:siz}
\end{figure}

In the third experiment, we fix the coherence tolerance $\alpha = 0.4$. Figure~\ref{fig:siz} shows the average interferer power and condition number when the number of randomized projections $L$ varies among $\left[10^0, 10^6\right]$. As shown in the figure, when the number of randomized projections is not sufficiently large, the sequence obtained in at least one iteration is suboptimal or even not feasible, which results in high values of the interferer power and condition number. Additionally, the necessary number of randomized projections decreases as the oversampling rate increases. When $L$ increases, we have decreasing interferer power, since the more randomized projections provide better interferer mitigation performance, as well as more stable condition numbers, which are bounded by the constraints.

In the final experiment, we test the interferer mitigation performance when signals containing off-grid frequencies are modulated by different sequences, including pseudorandom sequences. We also test the sequences obtained from RSDPR based on Fourier and Slepian bases, which we called Single Fourier and Single Slepian sequences, respectively. We compare those sequence sets with the sequence sets obtained from the proposed method shown in MB-RSDPR using both bases, which are denoted by Multiple Fourier and Multiple Slepian. Additionally, we include the MIMO SHAPE algorithm, which aims to extend the SHAPE algorithm~\cite{akay2017extending}, described in~\cite{rowe2014spectrally}, to obtain a set of complex-valued sequences that simultaneously approach the desired spectrum magnitude and satisfy the envelope constraint. With the similar methods shown in~\cite{Mo2015Design-of-Spect, Mo:2018a}, we modify the projection step in MIMO SHAPE algorithm onto the feasible set of unimodular sequences to return a set of binary sequences based on the Fourier and Slepian bases, which are denoted by SHAPE Fourier and SHAPE Slepian. While all sequences (except pseudorandom) are designed to block the frequency range $\left[f _c - 1 / RN, f _c + 1 / RN\right]$, the input signal represents a single interferer expressed by the complex exponential vector
\begin{align*}
\cF \left( f_c + \frac{d}{RN} \right) = \frac{1}{\sqrt{RN}} { \left[ 1, \dots, e ^{ j 2 \pi (RN - 1) \left( f _c + \frac{d}{N} \right) } \right] } ^T,
\end{align*}
where $d$ denotes the frequency offset to the center frequency $f _c$ of the interferer band. The output performance is then measured by the modulation gain, which is defined as the power of the modulated signals normalized by the sequence power, i.e., $\left\| \ms \cF \left( f _c + d / RN \right) \right\| _2 ^2 / RN$.

\begin{figure}[t]
\centering
\includegraphics[width = 0.45\textwidth]{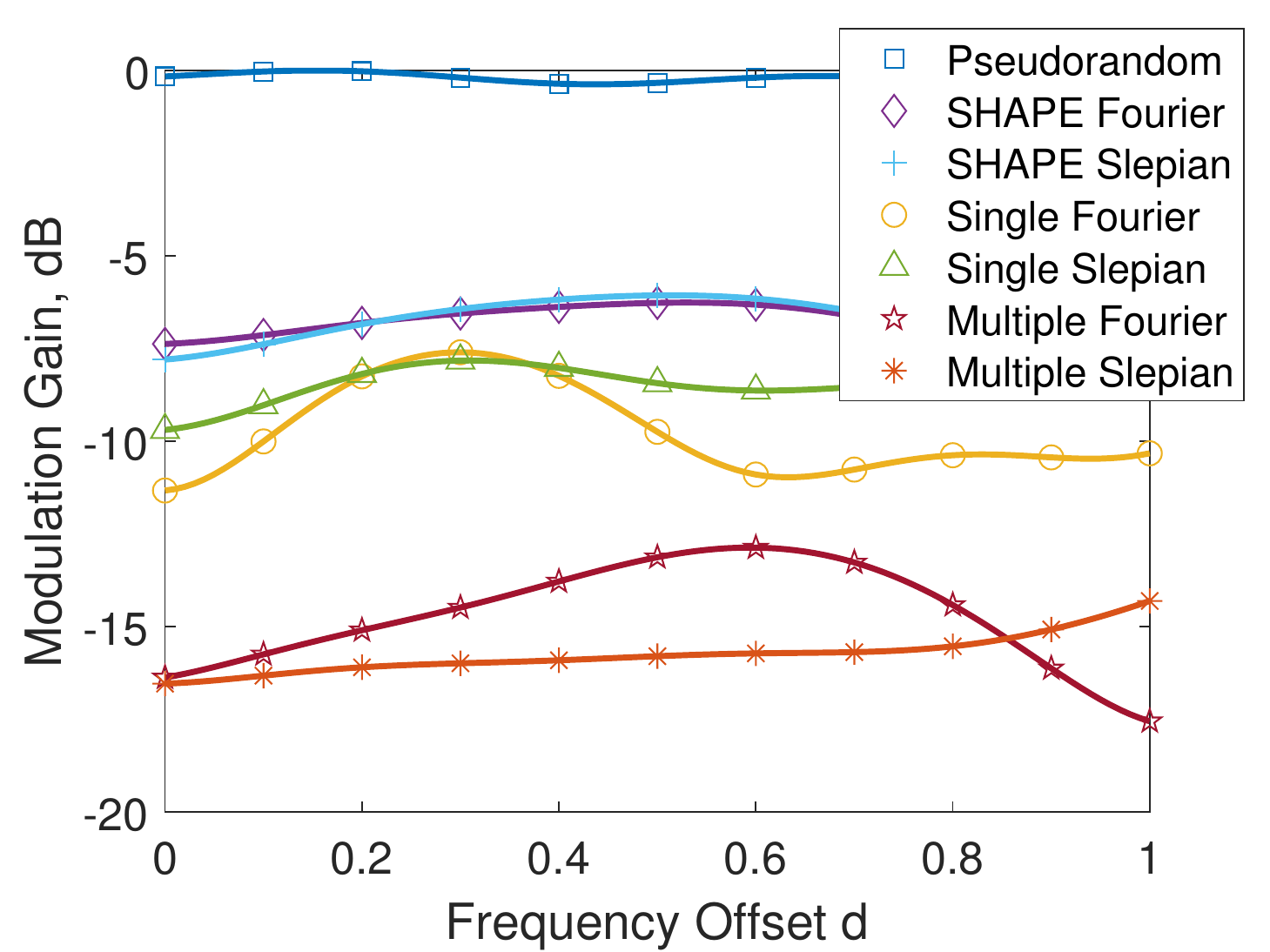}
\caption{Average modulation gain within the rejection band as a function of the frequency offset $d$ from the band center. Lower modulation game is better. The sequences designed based on the Fourier basis have small gain at on-grid frequencies, but large gain at frequencies far away to any on-grid frequency. The sequences designed based on the Slepian basis have almost the same gain across most frequencies.}\label{fig:off}
\end{figure}

Figure~\ref{fig:off} shows the average modulation gain for the interferer band over $100$ independently generated sequence sets as a function of the frequency offset $d$. The lower that the modulation gain is, the better performance that the sequence set has. Since the pseudorandom sequences have a flat spectrum, the modulation gain of the pseudorandom sequences is almost $0~\mathrm{dB}$ at all frequency offsets. The sequences from MC-RSDPR when the Fourier basis is used have good rejection for the on-grid frequencies (i.e., the frequency offset is $0$ or $1$). However, the rejection performance decreases as the interferer frequency moves farther from the on-grid frequencies. By contrast, the sequences based on the Slepian basis provide good rejection for most frequencies, both on-grid and off-grid, except for those that are close to the edge of the interferer bands.

\section{Conclusions and Future Work}\label{section:summary}

In this paper, we proposed an algorithm to design a set of binary sequences for multi-branch modulation receiver architecture that provides a notch for interferer bands while the message can be recovered from the modulated signals. The sequence design for each branch can be written as a QCQP that minimizes the sequence power in the interferer band with a set of constraints that enforce approximate mutual orthogonality among the sequences in the set. We provide an analysis of the difficulty of searching orthogonal binary sequence sets for stable recovery and highlight the necessity of oversampling in sequence design. We numerically showed that the performance of the designed sequence sets increases as the number of randomized projections increases. Our experiments show that the sequence sets based on the Slepian basis are more robust to the interferer component with off-grid frequency than those based on the Fourier basis.

Many questions remain open on the analysis of our algorithm. An interesting aspect would be to consider the statistics of the binary sequences. Since the binary sequences are obtained by quantizing the random vectors drawn from the solution of SDP relaxation, it is possible to derive a distribution for the entries of those binary sequences as a function of a multivariable Gaussian distribution. Thus a measurement similar to~(\ref{eq:SSBS_feasibility}) about the mutual orthogonality among the binary sequences could be obtained.

Furthermore, one could pursue additional optimization methods to improve sequence set design performance. For example, one could consider changes to the objective function and the constraints (e.g., switching the two) such that the optimization searches for sequence sets with minimal orthogonality while the sequence power in the interferer band is bounded. This may be beneficial by allowing for sequence sets with better recovery performance. Furthermore, there may exist more efficient approaches to characterize the sequence set orthogonality than coherence that we employed. For example, matrix preconditioning has been used to reduce the condition number of a matrix involved in a linear system iterative solver~\cite{SAAD20001, BENZI2002418}. However, an integration of matrix preconditioning into the optimization in our approaches is nontrivial. Alternatively, the cross-energy spectral density in ~\cite{akay2017extending}, which is defined as the Fourier transform of cross-correlation of sequences, has the potential to obtain sequences with low coherence. Additionally, the iterative method can propagate the error from one sequence to another sequence due to the constraints since each sequence is approximately optimal. It is promising to extend the randomized approaches to leverage matrix optimization such that the entire sequence set is obtained simultaneously.

\section*{Acknowledgement}

We thank Dennis Goeckel, Robert Jackson, Joseph Bardin, and Mohammad Ghadiri Sadrabadi for helpful comments to the authors during the completion of this research.

\end{document}